\documentclass[11pt,a4paper]{article}
\usepackage{amsmath, amssymb, amsthm, a4wide}

\newtheorem{thm}{\noindent Theorem}[section]
\newtheorem{lem}[thm]{\noindent Lemma}
\newtheorem{cor}[thm]{\noindent Corollary}
\newtheorem{prop}[thm]{\noindent Proposition}
\theoremstyle{remark}
\newtheorem{remark}{\noindent Remark}

\newcommand{\rd}{\mathrm{d}} 
\newcommand{\re}{\mathrm{e}} 
\newcommand{\rE}{\mathbb{E}} 

\skip\footins=17.07pt

\newcommand{\addressone}[1]{\footnote{\hspace*{-14pt}
      $^{*}\,$Postal address: #1}\par}

\def\cov{\mathop{\rm cov}\nolimits}    
\def\dens{\mathop{\rm dens}\nolimits}  
\def\vol{\mathop{\rm vol}\nolimits}    
\def\card{\mathop{\rm card}\nolimits}  
\def\ts{\hspace{0.5pt}}

\title{Absence of singular continuous diffraction\\
for discrete multi-component particle models}
\author{Michael Baake$^{*}$, Natali Zint$^{*}$}
\date{}

\begin{document}

\maketitle

\addressone{Fakult\"at f\"ur Mathematik, Universit\"at Bielefeld, 
 Postfach 100131, 33501 Bielefeld, Germany. 
 E-mail: \{mbaake,nzint\}@math.uni-bielefeld.de}

\begin{abstract}
  Particle models with finitely many types of particles are
  considered, both on $\mathbb{Z}^d$ and on discrete point sets of
  finite local complexity. Such sets include many standard examples of
  aperiodic order such as model sets or certain substitution systems.
  The particle gas is defined by an interaction potential and a
  corresponding Gibbs measure.  Under some reasonable conditions on
  the underlying point set and the potential, we show that the
  corresponding diffraction measure almost surely exists and consists
  of a pure point part and an absolutely continuous part with
  continuous density. In particular, no singular continuous part is
  present.
\end{abstract}

\bigskip




\section{Introduction}

Periodic systems in general, and lattice systems in particular, are at
the basis of crystallographic structure determination from (kinematic)
diffraction data. The diffraction spectrum (or measure) of a
crystallographic (or fully periodic) system in $\mathbb{R}^d$, with
lattice of periods $\varGamma$, is a positive pure point measure,
concentrated on the dual lattice $\varGamma^*$, the latter defined as
$\varGamma^* = \{ z \in \mathbb{R}^d \, | \, z \cdot y \in \mathbb{Z},
\forall y \in \varGamma\}$.

In reality, however, crystals are never perfect, and deviations from
full periodicity are abundant.\ One interesting class of extensions to
consider is that of the so-called lattice gases, where different types
of atoms (or one type and ``zeronium'') occupy the lattice sites
according to some random process, e.g., one described by a Gibbs
measure. The relevance of Gibbs measures, in turn, stems from the
assumption that the structure under consideration is in thermal
equilibrium. Clearly, this need not always be the case, but we do not
consider non-equilibrium systems here.

In two recent publications \cite{baak04,hoef00}, it was shown that
certain binary lattice gases, such as those based upon a ferromagnetic
Ising model with short-range interaction, additionally inherit an
absolutely continuous spectral component, but no singular continuous
one. This finding is in agreement with the working experience in
crystallography (compare \cite{cowl95,guin94,welb04}), and the natural
question arises whether this situation is more general.

The answer to this question is affirmative, and it is the aim of this
contribution to extend the results of \cite{baak04} to increase the
evidence for the rather natural conjecture that non-trivial singular
continuous diffraction spectra are the exception, at least as the
result of stochastic deviations from systems with a strongly ordered
ground state, which are of special interest (see \cite{vane92}).

Starting with a brief summary of the concept of finite local
complexity (FLC), we show that, under some condition on the potential
(such as finite or short range) and sufficiently high temperature,
there is no singular continuous diffraction component for a lattice
gas with finitely many types of particles. Our aim here is a
simultaneously more general and more transparent approach in
comparison to \cite{baak00,baak04}, where only binary systems were
considered. Also, the focus is different both from \cite{kuel03s},
where an approach via finite subsystems is used for the interaction
free case, and from \cite{kuel03}, where a large deviation estimate
for the diffraction of a finite point set is derived.

{}Finally, after various explicit examples in Section~\ref{s:gen_res},
we prove an analogous result for a certain class of FLC sets in
Section~\ref{extend}.


\section{FLC sets} \label{s:flc}

In what follows, we consider only point sets $\varGamma \subset
\mathbb{R}^d$ that are locally finite, i.e., the intersection
$\varGamma \cap K$ with any compact $K \subset \mathbb{R}^d$ is a
finite set.  Such a $\varGamma$ has \emph{finite local complexity}
(FLC), if, for any compact set $K \subset \mathbb{R}^d$, there are
only \emph{finitely} many clusters $\varGamma \cap (t+K)$, $t \in
\mathbb{R}^d$, when counted up to translations.  By a characterisation
of Schlottmann \cite{schl00}, a point set $\varGamma \subset
\mathbb{R}^d$ is FLC if and only if $\varGamma - \varGamma := \{ x-y
\, | \, x,y \in \varGamma\}$ is discrete and closed (or locally
finite).  This characterisation automatically includes the local
finiteness of $\varGamma$.  Clearly, FLC sets $\varGamma$ are also
uniformly discrete, i.e., there is a minimal distance between points
of $\varGamma$, as $0$ is an isolated point of $\varGamma -
\varGamma$.

The FLC property of $\varGamma$ has the consequence that the orbit $\{
t + \varGamma \, | \, t \in \mathbb{R}^d\}$ is precompact in the local
topology (LT), where two point sets are close if, after a small
translation, they agree on a large ball around $0 \in \mathbb{R}^d$.
Consequently,
\[
    \mathbb{X}(\varGamma):= \overline{\{ t + \varGamma \, | \, 
    t \in \mathbb{R}^d\}}^{\text{LT}}
\]
is compact, and $(\mathbb{X}(\varGamma),\mathbb{R}^d)$ forms a
topological dynamical system, on which one can define translation
invariant probability measures, see \cite{schl00} for details. The
latter, at this point, need neither be unique nor ergodic.  This can
be seen by combining $2 \mathbb{Z}$ with an arbitrary (e.g., random)
subset of $2 \mathbb{Z} + 1$, which always produces an FLC set (and
even a Meyer set, compare \cite{mood97}). When using a standard
Bernoulli process (e.g., coin tossing) to decide on the occupation of
the sites from $2 \mathbb{Z} + 1$, the hull $\mathbb{X}(\varGamma)$
almost surely contains the set $2 \mathbb{Z}$ as well as the set
$\mathbb{Z}$ (since there are $0$- and $1$-sequences of arbitrary
length in almost all realisations of the Bernoulli process).
Therefore, different probability measures on $\mathbb{X}(\varGamma)$
clearly exist.

{}From now on, we restrict ourselves to FLC sets $\varGamma \subset
\mathbb{R}^d$ such that $(\mathbb{X}(\varGamma),\mathbb{R}^d)$ is
uniquely ergodic or even strictly ergodic, i.e., uniquely ergodic and
minimal.  In this setting, unique ergodicity is equivalent to
$\varGamma$ having uniform cluster frequencies (UCF), and minimality
to $\varGamma$ being repetitive, see \cite{schl00} for details. This
class contains lattices, but also systems obtained by primitive
substitution rules \cite{laga03} as well as regular generic model sets
(also called cut and project sets) \cite{mood00}, hence two of the
most important generalisations of lattices.  Due to unique ergodicity,
we then have a unique invariant probability measure $\vartheta$ on
$\mathbb{X}(\varGamma)$, and
$(\mathbb{X}(\varGamma),\mathbb{R}^d,\vartheta)$ is both a topological
and a measurable dynamical system.

\subsection{A result from Fourier analysis}

Below, we need the following result from Fourier analysis concerning
FLC sets.

\begin{prop}\label{p:fourier_converges}
  Let\/ $\varGamma \subset \mathbb{R}^d$ be an FLC set, and consider
  the complex measure
\[
\nu := \sum_{z \in \varGamma -\varGamma} g(z)\; \delta_z \; .
\]
If\/ $\sum_{z \in \varGamma -\varGamma} \lvert g(z) \rvert < \infty$,
$\nu$ is a finite measure, and\/ the Fourier transform $\widehat{\nu}$
is a bounded, uniformly continuous function that defines an absolutely
continuous measure on\/ $\mathbb{R}^d$. Moreover, $\widehat{\nu}$ is a
generalised Fourier series,
\[
   \widehat{\nu}(k)=\sum_{z \in \varGamma -\varGamma} g(z)\; 
   \re^{-2\pi i k z}\; ,
\]
which converges uniformly and absolutely.
\end{prop}

\begin{proof}
  Recall that a regular Borel measure $\nu$ is finite when $\Vert \nu
  \Vert = \vert \nu \vert$ ($\mathbb{R}^d$) $<\infty$. Our $\nu$ is a
  pure point measure because $\varGamma -\varGamma$ is discrete and
  closed by the FLC property, so that $\Vert \nu \Vert =\sum_{z \in
    \varGamma -\varGamma} \lvert g(z) \rvert < \infty$ implies that
  $\nu$ is finite.

  By the general properties of the Fourier (or Fourier-Stieltjes)
  transform, compare \cite[Theorem 1.3.3]{rudi90}, $\widehat{\nu}$ is
  then a bounded and uniformly continuous function on $\mathbb{R}^d$.
  It is thus locally integrable, and hence defines an absolutely
  continuous measure on $\mathbb{R}^d$ by the Radon-Nikodym theorem.

  The absolute (and hence also uniform) convergence of the generalised
  Fourier series follows from $\Vert \nu \Vert < \infty$ together with
  $\widehat{\delta}_z(k)=\re^{-2\pi i k z}$.
\end{proof}

If $\varGamma -\varGamma$ is not only locally finite, but also
uniformly discrete, one has the following explicit sufficient
condition, which gives an easy access to the result of
Proposition~\ref{p:fourier_converges}.

\begin{lem}\label{c:condition}
  Let\/ $\varGamma \subset \mathbb{R}^d$ be an FLC set with\/
  $\varGamma -\varGamma$ uniformly discrete. If\/ $g(z)$ is a complex
  function with
\[
g(z) = \mathcal{O} \left( \vert z \vert^{-d-\varepsilon} \right)
\]
for\/ $\vert z \vert \to \infty$ and some $\varepsilon > 0$, the
measure\/ $\nu = \sum_{z \in \varGamma -\varGamma} g(z)\, \delta_z$ is
a finite measure, and Proposition~$\ref{p:fourier_converges}$ applies.
\end{lem}

\begin{proof}
  $\varGamma -\varGamma$ uniformly discrete means that there is some
  $r>0$ so that any translate of the open ball $B_r(0)$ contains at
  most one point of $\varGamma -\varGamma$. Select a suitable lattice
  $L \subset \mathbb{R}^d$ so that $\mathbb{R}^d=L+B_r(0)$. For a
  lattice in $\mathbb{R}^d$, the sum
\[
    \sum_{t \in L} \frac{1}{\vert t \vert^{d+\varepsilon}}
\]
is convergent. By standard arguments, one can now see that the sum
$\sum_{z \in \varGamma -\varGamma} g(z)$ is absolutely convergent as a
consequence of the assumed asymptotic properties. This gives $\Vert
\nu \Vert < \infty$.
\end{proof}

\subsection{A result from the theory of Gibbs measures}

Let\/ $\varGamma \subset \mathbb{R}^d$ be an FLC set with\/ $\varGamma
-\varGamma$ uniformly discrete.  Put $\mathcal{S}:=\{F \subset
\varGamma \, | \, 0<|F|<\infty \}$ (the finite subsets of $\varGamma$,
with $|F|:=\card(F)$) and consider an interaction potential
$(\varPhi_S)_{S \in \mathcal{S}}$ such that, for all $F \in
\mathcal{S}$ and all $\omega \in \{c_1, \dotsc , c_n\}^{\varGamma}$
with $c_i \in \mathbb{C}$, the total energy
\[
   \mathcal{H}_{F}^{\varPhi}(\omega)=\sum_{S \in 
   \mathcal{S},\, S \cap F \neq \varnothing} \varPhi_S(\omega)
\]
of $\omega$ in $F$ for $\varPhi$ exists. Let $\bar{F}= \varGamma
\backslash F$ be the complement of $F$ and let $ab$ with $a \in \{c_1,
\dotsc , c_n\}^{F}$ and $b \in \{c_1, \dotsc , c_n\}^{\bar{F}}$ denote
a combined configuration on $\varGamma$.  In particular, we use this
for $b=\omega_{|_{\bar{F}}}$.  The partition function for a finite
subsystem $F$ is given by
\[
Z_{F}^{\varPhi}(\omega)= \sum_{a \in \{c_1, \dotsc , c_n\}^{F}}
\re^{-\mathcal{H}_{F}^{\varPhi} (a\omega_{|_{\bar{F}}})}\; ,
\]
and the probability measure
\[
A \mapsto \mu^{}_{F} (A\, | \, \omega)\; = \;
\frac{1}{Z_{F}^{\varPhi}(\omega)} \sum_{a:\, a\omega_{|_{\bar{F}}} \in A}
\re^{-\mathcal{H}_{F}^{\varPhi}(a\omega_{|_{\bar{F}}})}\; ,
\]
for all $A \in \mathcal{A}$ (the product $\sigma$-algebra on $\{c_1,
\dotsc , c_n\}^{\varGamma}$), is called the Gibbs distribution in $F$
with boundary condition $\omega_{|_{\bar{F}}}$ and interaction
potential $\varPhi$, and defines the associated Gibbs measure $\mu$,
see \cite{geor88} for details.

We now show that, on one essential condition, the covariance of
$\Omega_x:\{c_1, \dotsc , c_n\}^{\varGamma} \rightarrow \{c_1, \dotsc
, c_n\}$, $\omega \mapsto \omega_x$ and $\overline{\Omega_y}:\{c_1,
\dotsc , c_n\}^{\varGamma} \rightarrow \{\overline{c_1}, \dotsc
,\overline{c_n}\}$, $\omega \mapsto \overline{\omega_y}$ is summable
in $y$, and that the sums are bounded for all $x$.

\begin{prop}\label{p:bounded_sums_of_covariances}
  Consider the FLC set\/ $\varGamma$ and any metric $\mathrm{d}(x,y)$.
  For\/ $S \in \mathcal{S}$, let\/ ${\rm diam}(S)$ denote the
  associated diameter of\/ $S$, and define
\[
   {\cal D} (\varPhi_S)=\sup_{\zeta,\lambda \in 
    \{c_1, \dotsc , c_n\}^{\varGamma}}
    \lvert \varPhi_S(\zeta)-\varPhi_S(\lambda)\rvert \, .
\]
On the condition that
\begin{equation}\label{e:central_condition}
    \sup_{u \in \varGamma} \sum_{u \in S \in \mathcal{S}} 
    \re^{{\rm diam}(S)} (|S|-1)\, {\cal D}(\varPhi_S)\; < \; 2 \; ,
\end{equation}
the corresponding Gibbs measure $\mu$ is unique. Moreover, for all $x
\in \varGamma$,
\[
   \sum_{y \in \varGamma} 
   \bigl| \cov_{\mu}(\Omega_x, \overline{\Omega_y}) \bigr|
   < \infty \, .
\]
\end{prop}

\begin{proof}
  Let $\lVert . \rVert$ be the variation norm.  Dobrushin's
  interaction matrix $C=(C_{uv})$ ($u,v \in \varGamma$) describes the
  dependence between the particles at different sites and is defined
  by
\begin{equation}\nonumber
  C_{uv}=\sup\left\{\lVert \mu_{\{u\}}(\cdot | 
    \zeta)-\mu_{\{u\}}( \cdot | \lambda) \rVert \, | \,
    \zeta=\lambda \text{ off } v\right\}\; .
\end{equation}
Thus, these elements quantify the largest variational distance between
the conditional distributions $\rd\mu_{\{u\}}(\omega_{|u}|\omega)$ for
a fixed site $u$ evaluated at two configurations which differ from
each other only at site $v$.

In the proof of \cite[Prop.~8.8]{geor88}, it was shown that
\begin{equation}\nonumber
  C_{uv} \leq \tfrac{1}{2} 
  \sum_{\{u,v\} \subset S \in \mathcal{S}} {\cal D}(\varPhi_S)\; .
\end{equation}
Therefore, due to \eqref{e:central_condition},
\begin{equation}\nonumber
\begin{split}
\alpha &:=\sup_u \sum_v \re^{\mathrm{d}(u,v)} 
  C_{uv} \leq \tfrac{1}{2} \sup_u \sum_{v \neq u} \sum_{S \supset \{u,v\}}
{\cal D}(\varPhi_S)\, \re^{\mathrm{d}(u,v)}\\
&\leq \tfrac{1}{2} \sup_u \sum_{u \in S \in \mathcal{S}} 
\re^{{\rm diam}(S)} (|S|-1)\, {\cal D}(\varPhi_S) < 1
\end{split}
\end{equation}
holds (which implies uniqueness of the Gibbs measure because $\sup_u
\sum_v C_{uv} < 1$), and we may apply \cite[Prop.~8.34]{geor88} which
states that
\begin{equation}\label{G:kovarianz}
| \cov_{\mu}(f,g)| \leq \tfrac{1}{4} 
\sum_{u,v \in \varGamma} {\cal D}_u(f) D_{uv} {\cal D}_v(g)\; ,
\end{equation}
where $f$ and $g$ are bounded quasilocal functions on $\{c_1, \dotsc ,
c_n\}^{\varGamma}$,
\[
   {\cal D}_u(f)=\sup \{ |f(\zeta)-f(\lambda)|
   \, | \, \zeta=\lambda \text{ off } u\}
\]
and $D_{uv}= \sum_{n=0}^{\infty} C_{uv}^n$. Recall that a measurable
function is called quasilocal when
\[
\lim_{F \in \mathcal{S}} \sup_{\substack{\zeta,\lambda \in
\{c_1, \dotsc , c_n\}^{\varGamma}\\ 
\zeta_{|_{F}}=\lambda_{|_{F}}}} \lvert f(\zeta) - f(\lambda) \rvert =0\; .
\]
Here, the notation $\lim_{F \in \mathcal{S}}$ means that the limit is
taken along sets, where more and more points are added, so that
$|F|\nearrow \infty$.

\smallskip Let $f(\omega)=\omega_x$ and
$g(\omega)=\overline{\omega_y}$. This results in
\[
{\cal D}_u(f)=\begin{cases}
  \sup_{i,j} |c_i-c_j|, & \text{if } u=x,\\
  0, & \text{otherwise,}
  \end{cases}
\]
and
\[
  {\cal D}_v(g)=\begin{cases}
   \sup_{i,j} |c_i-c_j|, & \text{if } v=y,\\
   0, & \text{otherwise.}
   \end{cases}
\]

Then, \eqref{G:kovarianz} implies
\[
   \sum_{y \in \varGamma} | \cov_{\mu}(\Omega_x, 
   \overline{\Omega_y})| \leq
   \sum_{y \in \varGamma} 
   \frac{\big( \sup_{i,j}|c_i-c_j|\big)^2}{4} 
        D_{xy} \leq
    \frac{\big( \sup_{i,j}|c_i-c_j|\big)^2}
                  {4(1-\alpha)}=c\, ,
\]
since \cite[Remark~8.26]{geor88} states that $\sum_{y \in \varGamma}
D_{xy} \leq 1/(1-\alpha)$.  Moreover, $c$ does not depend on $x$,
wherefore we get the result.
\end{proof}

\section{Diffraction theory}

The diffraction measure of a solid describes the outcome of a
kinematic diffraction experiment, e.g., by $X$-rays. It is the Fourier
transform of the autocorrelation measure of the solid, see
\cite{hof95} for the underlying theory in terms of translation bounded
(complex) measures and \cite{cowl95,guin94} for general background on
the physics of diffraction.

\subsection{Diffraction of a perfect lattice}

If we start with a perfect lattice $\varGamma$ in Euclidean space
(i.e., a co-compact discrete subgroup of $\mathbb{R}^d$) containing
$0$, its Dirac comb $\delta_{\varGamma}:=\sum_{z \in \varGamma}
\delta_z$ is a model for the atomic positions (of a mono-atomic
crystal, say). In this case, the autocorrelation is simply given by
\[
\gamma= \sum_{z \in \varGamma} \eta(z) \delta_z
\]
with the autocorrelation coefficients
\[
  \eta(z):= \lim_{r \to \infty} \frac{1}{\vol(B_r)} 
  \card\big(\varGamma_{r} \cap (z+\varGamma_{r})\big)\;,
\]
where $\varGamma_{r}:=\varGamma \cap B_r(0)$ with $B_r(0)$ the open
ball of radius $r$ centred at $0$.  Clearly, $\eta(z)=
\dens(\varGamma)$ for all $z \in \varGamma$, so that $\gamma =
\dens(\varGamma) \cdot \delta_{\varGamma}$.  Observing Poisson's
summation formula for Dirac combs \cite[Eq.~VII.7.4]{schw98},
\begin{equation} \label{e:PSF}
\widehat{\delta}_{\varGamma}=\dens(\varGamma) \cdot \delta_{\varGamma^*}\; ,
\end{equation}
with the dual lattice $\varGamma^*$, one obtains the well-known formula
\begin{equation}
\widehat{\gamma}=(\dens(\varGamma))^2 \cdot \delta_{\varGamma^*}
\end{equation}
for the diffraction from a perfect lattice.

This can easily be generalised to any measure of the form
\[
\nu= \rho * \delta_{\varGamma}
\]
with $\rho$ some finite measure, which is well-defined, compare
\cite[Prop.\ 1.13]{berg75}.  The diffraction then gives
\[
  \widehat{\gamma}_{\nu}=\lvert 
  \widehat{\rho} \ts \rvert^2 \cdot (\dens(\varGamma))^2 \cdot
  \delta_{\varGamma^*}
\]
by an application of the convolution theorem. The finite measure
$\rho$ can accomodate the distribution of finitely many possibly
different atoms over the unit cell of $\varGamma$ as well as
characteristic profiles of the atoms. The result shows up as the
continuous modulation factor $\lvert \widehat{\rho} \ts \rvert^2$ in
the diffraction measure, see \cite{cowl95} for various applications.

\subsection{Diffraction of lattice gases} \label{ss:lattice_gases}

We now turn to the rather general situation of a lattice $\varGamma$
occupied with $n \in \mathbb{N}$ different types of scatterers with
scattering strengths $\{c_1, \dotsc, c_n\}$, $c_i \in \mathbb{C}$,
subject to some stochastic process in equilibrium. Let $(H_x)_{x \in
  \varGamma}$ be a family of random variables describing the
scattering strengths at the lattice positions. Since each $H_x$ takes
one of finitely many finite values, $h_x \in \{c_1, \dotsc, c_n\}$,
the second moments of the random variables exist.  Moreover, we assume
that the family of random variables is controlled by a translation
invariant Gibbs measure $\mu$. Furthermore, we suppose that $\mu$ is
ergodic. The natural autocorrelation, if it exists, is given by
\[
\gamma^{(H)}= \sum_{z \in \varGamma} \eta^{(H)}(z) \; \delta_z \; ,
\]
where, with $\varGamma_r=\varGamma \cap B_r$,
\[
  \eta^{(H)}(z):= \lim_{r \to \infty} \frac{1}{\vol(B_r)} 
  \sum_{x \in \varGamma_r} H_x \overline{H_{x-z}}\;.
\]

\begin{lem}\label{l:lemma1}
  With the above-mentioned assumptions, the equation
\[
  \gamma^{(H)}=\lvert \rE_{\mu}(H_0) \rvert^2 
  \; \gamma + \dens(\varGamma)\; \sum_{z \in \varGamma}
\cov_{\mu}(H_0,\overline{H_{-z}}) \; \delta_z
\]
holds $\mu$-a.s., where $\gamma$ is the autocorrelation of the fully
occupied lattice and one has the relation
$\cov_{\mu}(H_0,\overline{H_{-z}})= \rE_{\mu}(H_0
\overline{H_{-z}})-\lvert \rE_{\mu}(H_0) \rvert^2$.
\end{lem}

\begin{proof}
  Let $T_x$ denote the shift map, i.e., let $T_x(H_0
  \overline{H_{-z}}):= H_x \overline{H_{x-z}}$.  Due to Birkhoff's
  ergodic theorem \cite[Chapter 2]{kell98}, one has $\mu$-a.s.:
\begin{equation}\nonumber
\begin{split}
  \eta^{(H)}(z)&= \lim_{r \to \infty} \frac{1}{\vol(B_r)} \sum_{x \in
    \varGamma_r} T_{x}(H_0 \overline{H_{-z}})=
  \lim_{r \to \infty} \frac{\card(\varGamma_r)}{\vol(B_r)}\; 
  \rE_{\mu}(H_0 \overline{H_{-z}})\\
  &= \dens(\varGamma) \; \rE_{\mu}(H_0 \overline{H_{-z}})\; .
\end{split}
\end{equation}
Then, by using the fact that
$\rE_{\mu}(\overline{H_{-z}})=\rE_{\mu}(\overline{H_0})$ for all $z
\in \varGamma$, one has
\begin{equation}\label{e:splitting}
\begin{aligned}
\gamma^{(H)} &= \sum_{z \in \varGamma} \dens(\varGamma) 
\; \rE_{\mu}(H_0\overline{H_{-z}}) \; \delta_z\\
&= \sum_{z \in \varGamma} \dens(\varGamma) \; 
\lvert \rE_{\mu}(H_0) \rvert^2 \; \delta_z +
\sum_{z \in \varGamma} \dens(\varGamma) \; 
\left( \rE_{\mu}(H_0\overline{H_{-z}}) -
\lvert \rE_{\mu}(H_0) \rvert^2 \right)\, \delta_z\\
&=  \lvert \rE_{\mu}(H_0) \rvert^2 \; \gamma + \dens(\varGamma) \;
\sum_{z \in \varGamma}\cov_{\mu}(H_0,\overline{H_{-z}}) \; \delta_z\; ,
\end{aligned}
\end{equation}
which establishes the claim.
\end{proof}
Note that the splitting in \eqref{e:splitting} makes sense in terms of
diffraction theory whenever the second term is well-behaved (e.g.,
when the covariance falls sufficiently rapidly).

\section{Special lattice gases} \label{s:gen_res}

In what follows, we analyse the diffraction spectra of lattice gases
as described in the previous section, subject to some local or
short-range interaction of stochastic nature. Here, we use $\varGamma
= \mathbb{Z}^d$ for simplicity, but the results do not really depend
on this particular choice of a lattice. In fact, they are robust under
non-singular affine transformations.  We show that, on one essential
condition, the diffraction spectrum of such a lattice gas model does
not contain a singular continuous part.

\begin{thm}\label{p:general_result}
On the condition that
\begin{equation}
\sup_{x \in \mathbb{Z}^d} \sum_{x \in S \in \mathcal{S}} 
\re^{{\rm diam}(S)} (|S|-1)\,
{\cal D}(\varPhi_S)\; < \; 2 \; ,
\end{equation}
the corresponding Gibbs measure $\mu$ is unique. Moreover, the
diffraction spectrum of the corresponding lattice gas model with\/ $n$
different types of particles $\mu$-a.s. exists, is
$\mathbb{Z}^d$-periodic and consists of a pure point part and an
absolutely continuous part with continuous density.  No singular
continuous part is present.
\end{thm}

\begin{proof}
  In the present case, as in Lemma~\ref{l:lemma1}, the autocorrelation
  measure is given by
\[
  \gamma^{(H)} = \sum_{x \in \mathbb{Z}^d} 
  \rE_{\mu}(H_0, \overline{H_{-x}})\, \delta_x
  =\lvert \rE_{\mu}(H_0) \rvert^2 \,\delta_{\mathbb{Z}^d}
  + \sum_{x \in \mathbb{Z}^d} \cov_{\mu}(H_0, 
  \overline{H_{-x}})\,\delta_x\; .
\]
This holds for $\mu$-almost all elements of our lattice gas ensemble
(respectively for $\mu$-almost all realisations of the corresponding
stochastic process).

The first part of the autocorrelation gives, under Fourier transform,
the pure point part of the diffraction measure, by means of the Poisson 
summation formula \eqref{e:PSF} for lattice Dirac combs (compare 
\cite{baak04} and references given there). This part is clearly
$\mathbb{Z}^d$-periodic.

Furthermore, due to Proposition~\ref{p:bounded_sums_of_covariances},
we have $\sum_{x \in \mathbb{Z}^d}\, \bigl| \cov_{\mu}(H_0,
\overline{H_{-x}})\bigr| <\infty$.  Therefore, by
Proposition~\ref{p:fourier_converges} with
$\varGamma-\varGamma=\mathbb{Z}^d$, the second part of the
autocorrelation, under explicit Fourier transform according to
Proposition~\ref{p:fourier_converges}, gives the absolutely continuous
part of the diffraction measure. It is again $\mathbb{Z}^d$-periodic
(compare \cite{MB,baak04} for a more general explanation of this
phenomenon), and our claim follows.
\end{proof}

Let us look at the implications of Theorem~\ref{p:general_result} for
some special models that are relevant to crystallographic
applications.

First, we consider finite range potentials $(\varPhi_S)_{S \in
  \mathcal{S}}=(\beta \varPhi^*_S)_{S \in \mathcal{S}}$ with $\beta:=
1/(k_B T)$ (the inverse temperature). This means that
\[
  \sup\, \big\{ \text{diam}(S)\, | \, S \in \mathcal{S},\, 
  \varPhi_S \neq 0 \big\} =R <\infty\; ,
\]
where $R$ denotes the range. For sufficiently small $\beta$, this
implies
\begin{equation}\nonumber
\sup_x \sum_{x \in S \in \mathcal{S}} \re^{{\rm diam}(S)} (|S|-1) 
\; {\cal D}(\varPhi_S)
\leq \re^R \sup_x \sum_{x \in S \in \mathcal{S}} 
(|S|-1)\; {\cal D}(\varPhi_S) \leq c\beta < 2\; ,
\end{equation}
where $c$ is a constant. Indeed, the finite number of $S \ni x$ with
$\varPhi_S \neq 0$ results in a finite sum. Moreover, because of the
finite range, there is a maximum of $|S|$ and ${\cal D}(\varPhi_S)$
for all $S$ with diam$(S) \leq R$.  Thus, for sufficiently high
temperatures, Condition \eqref{e:central_condition} holds and,
according to Theorem~\ref{p:general_result}, the diffraction spectrum
of such a model contains no singular continuous part for such
temperatures. This is most likely true in more generality, but does
not follow from our simplified approach.

\begin{remark}
  A well-known model of this type is the Potts model, where only pair
  interactions of nearest neighbours occur. This includes the classic
  ferromagnetic Ising model, where absence of singular continuous
  diffraction is already known for all temperatures, including the
  critical point \cite{baak00,baak04}.
\end{remark}

Now, we consider potentials with exponentially or algebraically
decaying pair interactions.  Thus, the interaction potential is given
by
\begin{equation}\nonumber
\varPhi_S(\omega)=
\begin{cases}
\beta \phi(\omega_{|_{x,y}})J(x-y), & \text{if $S=\{x,y\}$,}\\
0, & \text{otherwise,}
\end{cases}
\end{equation}
with $\phi:\{c_1, \dotsc , c_n\} \times \{c_1, \dotsc , c_n\}
\rightarrow \mathbb{C}$, and we suppose that we have either
\begin{equation}\label{e:exponentially}
\sum_{x \in \mathbb{Z}^d} \re^{t \lVert x  \rVert} |J(x)|<\infty
\end{equation}
or
\begin{equation}\label{e:algebraically}
\sum_{x \in \mathbb{Z}^d} \lVert x  \rVert^p |J(x)|<\infty
\end{equation}
for some positive constants $t$ and $p>1$.  Then, for sufficiently
high temperatures, the diffraction spectrum for such a model contains
no singular continuous part.

\bigskip

\noindent {\bf Argument:}
Consider the metric $ \mathrm{d} (x,y) := t \lVert x - y \rVert
\wedge \lfloor p \rfloor \log (1+ \lVert x - y \rVert)$ with some
constants $t>0$ and $p>1$, where $a \wedge b$ means the minimum
of $a$ and $b$. It is not difficult to check that this is indeed
a metric. Then, \eqref{e:exponentially} results, for 
sufficiently high temperatures, in
\[
  \sup_x \sum_{x \in S \in \mathcal{S}} \re^{{\rm diam}(S)} 
  (|S|-1)\, {\cal D}(\varPhi_S)
  \leq \beta \; \sup(|\phi(\zeta_{|_{x,0}})-\phi(\lambda_{|_{x,0}})|)
  \sum_{x \in \mathbb{Z}^d} \re^{t \lVert x  \rVert} |J(x)| <2
\]
and, again for sufficiently high temperatures,
\eqref{e:algebraically} results in the estimate
\[
  \sup_x \sum_{x \in S \in \mathcal{S}} \re^{{\rm diam}(S)} 
  (|S|-1) {\cal D}(\varPhi_S)
  \leq c \beta \; \sup(|\phi(\zeta_{|_{x,0}})-\phi(\lambda_{|_{x,0}})|)
  \sum_{x \in \mathbb{Z}^d}\lVert x  \rVert^p |J(x)|< 2\; .
\]
Note that, in addition to ${\rm diam}(S) \leq t \lVert x - y \rVert$,
one has ${\rm diam}(S) \leq \lfloor p \rfloor \log (1+ \lVert x - y \rVert)$
in this metric, which implies the estimate 
\[
   \re^{{\rm diam}(S)} \leq
   \left( 1 + \lVert x  \rVert \right)^{\lfloor p \rfloor} =
   \sum_{k=0}^{\lfloor p \rfloor} \binom{\lfloor p \rfloor}{k}
   \lVert x  \rVert^{\lfloor p \rfloor} \leq
   \sum_{k=0}^{\lfloor p \rfloor} \binom{\lfloor p 
   \rfloor}{k} \lVert x  \rVert^p =
   c \, \lVert x  \rVert^p .
\]
Thus, applying Theorem~\ref{p:general_result} once more, we obtain 
the absence of singular continuous diffraction for all sufficiently high 
temperatures.
\qed

\section{Extension to more general FLC sets} \label{extend}

The result of Theorem~\ref{p:general_result} is inherently robust and
one would expect analogous results beyond the lattice case. Of
particular interest (e.g., in the mathematical theory of
quasicrystals) is the general class of point sets of finite local
complexity, as introduced in Section~\ref{s:flc}.

\subsection{Diffraction without disorder}

In our setting with $\varGamma$ an FLC set and
$(\mathbb{X}(\varGamma),\mathbb{R}^d)$ uniquely ergodic, each element
$\varGamma' \in \mathbb{X}(\varGamma)$ is itself an FLC set. (Note
that, due to the fact that $\varGamma - \varGamma$ is locally finite,
any element of $\mathbb{X}(\varGamma)$ can at most possess clusters
that also occur in $\varGamma$.) The corresponding Dirac comb
$\delta_{\varGamma'}$ is then a translation bounded (and thus
tempered) measure whose autocorrelation
\[
\gamma^{}_{\varGamma'} := \lim_{r \to \infty} \frac{1}{\vol(B_r)}\,
\delta_{\varGamma' \cap B_r} * (\delta_{\varGamma' \cap B_r})^{\sim}
\]
exists, where $\tilde{\nu}$ is the measure defined by $\tilde{\nu}(g)
=\overline{\nu(\tilde{g})}$ with $\tilde{g}(x):=\overline{g(-x)}$, see
\cite{hof95,schl00} for details.

\begin{prop}\label{l:same_autocorr}
  Let\/ $\varGamma$ be an FLC set such that\/
  $(\mathbb{X}(\varGamma),\mathbb{R}^d)$ is uniquely ergodic. Then,
  each\/ $\varGamma' \in \mathbb{X}(\varGamma)$ has the same
  autocorrelation\/ $\gamma$.  The latter is a positive and positive
  definite pure point measure that can be written as
\[
\gamma = \sum_{z \in \varGamma - \varGamma} \eta(z) \delta_z\; ,
\]
with the autocorrelation coefficients
\[
\eta(z) := \lim_{r \to \infty} \frac{1}{\vol(B_r)}
\card\big(\varGamma_r \cap (z + \varGamma_r)\big)\; .
\]
\end{prop} 

\begin{proof}
  By definition of $\mathbb{X}(\varGamma)$, each $\varGamma' \in
  \mathbb{X}(\varGamma)$ is the limit of some sequence of the form
  $(t_n + \varGamma)$ in the local topology. Clearly, for any $t \in
  \mathbb{R}^d$, $\varGamma$ and $t + \varGamma$ possess the same
  autocorrelation, i.e., $\gamma^{}_{t +
    \varGamma}=\gamma^{}_{\varGamma}$. Moreover, unique ergodicity
  implies that
\[
\gamma_{t + \varGamma}^{(r)} := \frac{1}{\vol(B_r)}
\delta_{(t+\varGamma) \cap B_r} * (\delta_{(t+\varGamma) \cap
  B_r})^{\sim} = \gamma_{\varGamma}^{(r)} + o(1)\; ,\quad \text{as}\;
r \to \infty\; ,
\]
where the $o(1)$ term is uniform in $t$ (due to UCF, see
Section~\ref{s:flc}).

Then, when considering $\gamma_{t_n + \varGamma}^{(r)}$, the two
limits $r \to \infty$ and $n \to \infty$ commute, which is tantamount
to saying that the mapping $\varGamma' \mapsto \gamma^{}_{\varGamma'}$
is continuous on $\mathbb{X}(\varGamma)$. Since $\gamma^{}_{t_n +
  \varGamma} \equiv \gamma^{}_{\varGamma}$, this means that
$\gamma^{}_{\varGamma'} \equiv \gamma^{}_{\varGamma} = \gamma$ on
$\mathbb{X}(\varGamma)$.

Translation boundedness, positivity and positive definiteness are
standard and clear by construction, while the explicit representation
of $\gamma$ as a pure point measure is a consequence of $\varGamma -
\varGamma$ being locally finite. The formula for $\eta(z)$ is a simple
calculation, the limit exists due to unique ergodicity,
compare~\cite{bale04} for details and for an alternative approach via
the unique ergodic measure $\vartheta$ on $\mathbb{X}(\varGamma)$.
\end{proof}

\begin{cor}\label{c:decomposition}
  Under the assumptions of Proposition~$\ref{l:same_autocorr}$, also
  the diffraction measure\/ $\widehat{\gamma}$ is the same for all
  elements of the hull\/ $\mathbb{X}(\varGamma)$. It is a positive,
  translation bounded measure on\/ $\mathbb{R}^d$, with a unique
  decomposition
\[
\widehat{\gamma} = \widehat{\gamma}_{\sf pp} + \widehat{\gamma}_{\sf
  sc} + \widehat{\gamma}_{\sf ac}
\]
into its pure point, singular continuous and absolutely continuous
parts, relative to Lebesgue measure as reference $($being the Haar
measure of\/ $\mathbb{R}^d)$. $\hspace*{\fill}\square$
\end{cor}

This generalises the setting of crystallographic sets, the latter
being examples of pure point diffractive systems. Other examples with
pure point diffraction are regular model sets, also called cut and
project sets \cite{mood00}. In general, however, other spectral types,
or mixtures, can occur \cite{baak00,hoef00,kaku72}.

\subsection{Influence of disorder}

Let us first look at an interaction-free particle gas on $\varGamma$,
where we continue to assume that $\varGamma$ is FLC with
$(\mathbb{X}(\varGamma),\mathbb{R}^d)$ uniquely ergodic. Let
$\vartheta$ be the unique translation invariant probability measure on
$\mathbb{X}(\varGamma)$. To describe the system, we consider an
i.i.d.\ family of random variables, labelled by the points of
$\varGamma$. For simplicity, we assume that each variable $H_x$ takes
a (complex) value $h_x \in \{c_1, \dotsc, c_n\}$, with attached
probabilities $p_1, \dotsc, p_n$.

In line with Corollary~\ref{c:decomposition}, let $\widehat{\gamma}$
denote the diffraction measure of $\varGamma$.  By standard arguments
\cite{baak98,baak00,kuel03s,kuel03}, one then obtains the following
result for the diffraction of the particle gas, where the $c_i$ serve
as weights that represent the scattering strength at the corresponding
point of $\varGamma$.

\begin{thm}
  Let\/ $\delta_{\varGamma}^{(H)}:= \sum_{x \in \varGamma} H_x \,
  \delta_x$ be a Bernoulli process on\/ $\varGamma$ with i.i.d.\
  random variables\/ $H_x$ that take values $h_x \in \{c_1, \dotsc,
  c_n\}$, according to a common probability vector $p$, and assume
  that $0 \in \varGamma$.

  Then, the autocorrelation\/ $\gamma^{(H)}$ almost surely exists and
  results in the diffraction measure
\[
\widehat{\gamma}^{(H)}=\lvert \rE_p(H_0) \rvert^2 \; \widehat{\gamma}
+ \dens(\varGamma)\; \left( \rE_p(|H_0|^2)-\lvert \rE_p(H_0) \rvert^2
\right)\; ,
\]
where the second term on the right hand side is a constant and hence a
contribution to the absolutely continuous part of\/
$\widehat{\gamma}^{(H)}$. $\hspace*{\fill}\square$
\end{thm}

The key in proving this result is to establish that the new
autocorrelation is still concentrated on $\varGamma-\varGamma$, with
coefficients
\begin{equation}\nonumber
\eta^{(H)}(z)=\lvert \rE_p(H_0) \rvert^2 \; \eta(z) + 
\dens(\varGamma)\; \left(
\rE_p(|H_0|^2)-\lvert \rE_p(H_0) \rvert^2 \right)\; \delta_{z,0}
\end{equation}
with $\rE_p(H_0)=\sum_{i=1}^n p_i c_i$ etc. Due to the i.i.d.\ nature,
the validity of this type of result can be extended to more general
systems, compare \cite{kuel03s,kuel03}.

\bigskip

We now begin to extend our analysis to more interesting types of
disorder. Instead of an i.i.d.\ family of random variables, we
consider a family of random variables which are controlled by a Gibbs
measure $\mu^{}_{\varGamma}$.  Moreover, we have the unique ergodic
measure $\vartheta$ on $\mathbb{X}(\varGamma)$ that is essentially
defined by the cluster frequencies $f_{\mathfrak{y}}$ for arbitrary
clusters $\mathfrak{y}$.  In the following, we consider the measures
\[
\delta_{\varGamma}=\sum_{x \in \varGamma} \delta_x\; ,\quad
\delta^{(H)}_{\varGamma}=\sum_{x \in \varGamma} H_x \delta_x \quad
\text{and} \quad
\rE_{\mu^{}_{\varGamma}}(\delta^{(H)}_{\varGamma})=\sum_{x \in
  \varGamma} \rE_{\mu^{}_{\varGamma}}(H_x) \delta_x\; .
\]
Moreover, we define $\mathbb{X}(\delta_{\varGamma})$ as the orbit
closure of $\delta_{\varGamma}$ in the vague topology. Note that there
is a topological conjugacy between 
$(\mathbb{X}(\varGamma),\mathbb{R}^d)$ 
and $(\mathbb{X}(\delta_{\varGamma}),\mathbb{R}^d)$, see 
\cite[Lemma~2]{bale04}.

\begin{thm}\label{t:diffraction_flc}
  Assume that $\delta_{\varGamma}$ has a pure point diffraction
  spectrum. On the three conditions that the map
  $\mathbb{X}(\delta_{\varGamma}) \ni \delta_{\varGamma'} \mapsto
  \rE_{\mu^{}_{\varGamma'}}(\delta^{(H)}_{\varGamma'})$ is continuous,
  that the natural autocorrelations of $\delta^{(H)}_{\varGamma}$ and
  $\rE_{\mu^{}_{\varGamma}}(\delta^{(H)}_{\varGamma})$ a.s.\ exist, and
  that the difference between the corresponding autocorrelation 
  coefficients is absolutely
  summable, the diffraction spectrum of the particle gas with $n$
  different types of particles consists of a pure point part and an
  absolutely continuous part with continuous density. No singular
  continuous part is present.
\end{thm}

\begin{proof}
  $(\mathbb{X}(\rE_{\mu^{}_{\varGamma}}(\delta^{(H)}_{\varGamma})),
  \mathbb{R}^d)$ is a factor of
  $(\mathbb{X}(\delta_{\varGamma}),\mathbb{R}^d)$.  Since
  $\delta_{\varGamma}$ has a pure point diffraction spectrum by
  assumption, this is inherited by
  $\rE_{\mu^{}_{\varGamma}}(\delta^{(H)}_{\varGamma})$ due to
  \cite[Prop.\ 1 and Thm.\ 2]{bale05}. Moreover, the difference
  between the autocorrelation coefficients is absolutely summable 
  and therefore
  only results in an absolutely continuous part of the diffraction
  spectrum.
\end{proof}

\begin{cor}\label{c:model_set_finite_range}
  In the case of a regular model set, finite range potential and
  sufficiently high temperatures, there is no singular continuous 
  part present in the diffraction spectrum of the particle gas
  on this set.
\end{cor}

\begin{proof}
  Continuity of the map $\mathbb{X}(\delta_{\varGamma}) \ni
  \delta_{\varGamma'} \mapsto
  \rE_{\mu^{}_{\varGamma'}}(\delta^{(H)}_{\varGamma'})$ is obvious due
  to the finite range potential.  Let $\mathfrak{Y}_{|z|}:=\{
  \text{clusters of radius } |z|+R \text{ around } 0\}$ be the set of
  all clusters with specified radius, where $R$ is the range of the
  potential.  A regular model set has uniform
  cluster frequencies $f_{\mathfrak{y}}$ (chosen such that
  $\sum_{\mathfrak{y} \in \mathfrak{Y}_{|z|}} f_{\mathfrak{y}} =1$),
  see \cite[Thm.~4.5]{schl00}. Together with the finite range
  potential this is the reason why we may use the strong law of large
  numbers and get
\[
\begin{split}
  \eta^{(H)}(z)&=\lim_{r \to \infty} \frac{1}{\vol(B_r)} \sum_{x \in
    \varGamma_r} H_x \overline{H_{x-z}} \\[1mm]
  & =\lim_{r \to \infty}
  \frac{\card(\varGamma_r)}{\vol(B_r)} \sum_{\mathfrak{y} \in
    \mathfrak{Y}_{|z|}} \frac{1}{\card(\varGamma_r)} \sum_{\substack{x
      \in \varGamma_r\\ (\varGamma-x) \cap B_{|z|+R}=\mathfrak{y}}}
  H_x\overline{H_{x-z}}\\
  &= \dens(\varGamma) \sum_{\mathfrak{y} \in \mathfrak{Y}_{|z|}}
  f_{\mathfrak{y}}\, \rE_{\mu^{}_{\varGamma}}(H_x
  \overline{H_{x-z}}|(\varGamma-x)
  \cap B_{|z|+R}=\mathfrak{y})\\
  &=\dens(\varGamma) \sum_{\mathfrak{y} \in \mathfrak{Y}_{|z|}}
  f_{\mathfrak{y}}\, \rE_{\mu^{}_{\varGamma}}(H_x|(\varGamma-x) \cap
  B_{|z|+R}=\mathfrak{y})\,
  \rE_{\mu^{}_{\varGamma}}(\overline{H_{x-z}}|(\varGamma-x) 
  \cap B_{|z|+R}=\mathfrak{y})\\
  &\quad + \dens(\varGamma) \sum_{\mathfrak{y} \in \mathfrak{Y}_{|z|}}
  f_{\mathfrak{y}}
  \cov_{\mu^{}_{\varGamma}}(H_x\overline{H_{x-z}}|(\varGamma-x) 
  \cap B_{|z|+R}=\mathfrak{y})\\
  &=\lim_{r \to \infty} \frac{1}{\vol(B_r)} \sum_{x \in \varGamma_r}
  \rE_{\mu^{}_{\varGamma}}(H_x)\,
  \rE_{\mu^{}_{\varGamma}}(\overline{H_{x-z}})\\
  &\quad +\dens(\varGamma) \sum_{\mathfrak{y} \in \mathfrak{Y}_{|z|}}
  f_{\mathfrak{y}}
  \cov_{\mu^{}_{\varGamma}}(H_x\overline{H_{x-z}}|(\varGamma-x) \cap
  B_{|z|+R}=\mathfrak{y})\; .
\end{split}
\]
Note that, in order to apply the strong law of large numbers, we first
have to split the last sum in the first line into a finite number of
sums over points with non-overlapping surroundings of radius $|z|+R$.
The law of large numbers is then applied to each sum separately, each
of which a.s.\ converges to the same limit, compare \cite{BBM,hof95,MR}
for related results.

In the end, the first term is the autocorrelation coefficient of the
averaged Dirac comb
$\rE_{\mu^{}_{\varGamma}}(\delta^{(H)}_{\varGamma})$ at $z$.  Along
the lines of Section~\ref{s:gen_res}, one can show that condition
\eqref{e:central_condition} holds.  Therefore, due to
Proposition~\ref{p:bounded_sums_of_covariances}, $\sum_{z \in
  \varGamma-\varGamma} \dens(\varGamma) \sum_{\mathfrak{y} \in
  \mathfrak{Y}_{|z|}} f_{\mathfrak{y}}\, \bigl|
\cov_{\mu^{}_{\varGamma}}(H_x\overline{H_{x-z}}|(\varGamma-x) \cap
B_{|z|+R}=\mathfrak{y})\bigr|$ is finite.  Moreover, the diffraction of a
regular model set is pure point (compare \cite[Thm.~4.5]{schl00}).
Therefore, from Theorem~\ref{t:diffraction_flc}, we obtain the absence
of a singular continuous part in the diffraction spectrum of the
particle gas.
\end{proof}

\begin{remark}
  In general, a singular continuous diffraction component of a
  particle gas on a graph in $\mathbb{R}^d$ seems possible due to two
  sources. On the one hand, it could be a peculiar long-range
  structure or irregularity of the graph, which is excluded by the
  model set assumption in our case.  On the other hand, it could be
  caused by some long-range order in the Gibbs or ground state
  measure, which is excluded by range condition together with the
  temperature assumption that leads to a Dobrushin uniqueness regime.
\end{remark}

\section{Outlook}

The results of this paper, taken together with those of
\cite{kuel03s,kuel03}, demonstrate that many results known from the
theory of lattice gases can be extended to more general types of
systems.  In particular, model sets behave in very much the same way
as crystals to the superposition of stochastic disorder. This is
also the case in related studies that employ point process
methods to derive explicit examples, compare \cite{BBM} and
references therein.

The conditions investigated here for the non-periodic systems are to
be considered as a first step. It is quite clear that the condition of
finite range in Cor.~\ref{c:model_set_finite_range} can be lifted, but
the proofs will then become technically more involved. This calls for
an alternative or modified point of view to simplify the approach.

In this paper, we have considered FLC sets together with configuration
spaces based on finite type spaces.  It is clear that one can now also
increase the generality towards countably many types per site, or even
to continuous type spaces. This certainly adds another layer of
technical complication, but we expect the principal results to remain
unchanged.

\bigskip

\noindent {\bf Acknowledgements:}
It is our pleasure to thank H.-O.\ Georgii, Yu.\ Kondratiev and M.\
R\"ockner for discussions and useful suggestions as well as K.\
Matzutt for useful hints on the manuscript. We also thank an anonymous
referee for a number of valuable comments and questions.  This work
was supported by the German Research Council (DFG), within the CRC 701.

\end{document}